\documentclass[12pt]{aastex}
\usepackage{epsfig} 
\begin{document}

\title{
    Experimental Limit to Interstellar $^{244}$Pu Abundance
}

\author{
M. Paul 	\altaffilmark{1},
A. Valenta	\altaffilmark{1,2}, 	
I. Ahmad 	\altaffilmark{3}, 
D. Berkovits	\altaffilmark{4}, 
C. Bordeanu	\altaffilmark{5},
S. Ghelberg	\altaffilmark{1,6},
Y. Hashimoto	\altaffilmark{7}, 
A. Hershkowitz	\altaffilmark{1,8}, 
S. Jiang	\altaffilmark{1,9}, 
T. Nakanishi	\altaffilmark{7,10} 
and 
K. Sakamoto	\altaffilmark{7,11}} 
\altaffiltext{1}
{Racah Institute of Physics, Hebrew University, Jerusalem, Israel 91904;
email: paul@vms.huji.ac.il}
\altaffiltext{2}
{and
Institut f\"{u}r Isotopenforschung und Kernphysik, VERA Laboratory, 
University of Vienna, A-1090 Vienna, Austria; email: valenta@ap.univie.ac.at}
\altaffiltext{3}
{Physics Division, Argonne National Laboratory, Argonne, IL 60439, U.S.A.;
email: ahmad@phy.anl.gov}
\altaffiltext{4}
{Soreq NRC, Yavne, Israel 81800; email: berkova@vms.huji.ac.il}
\altaffiltext{5}
{Particle Physics Dept., Weizmann Institute of Science, Rehovot, Israel 76100;
bordeanu@wicc.weizmann.ac.il}
\altaffiltext{6}
{email: stelian@vms.huji.ac.il}
\altaffiltext{7}
{Dept. of Chemistry, Faculty of Science, Kanazawa University,
Kakumamachi, Kanazawa 920-1192, Japan}
\altaffiltext{8}
{email: aviher@vms.huji.ac.il}
\altaffiltext{9}
{email: jiang@vms.huji.ac.il}
\altaffiltext{10}
{email: nakanisi@cacheibm.s.kanazawa-u.ac.jp}
\altaffiltext{11}
{email: kohsakamoto@par.odn.ne.jp}

\begin{abstract}
Short-lived nuclides, now extinct in the solar system, are expected
to be present in interstellar matter (ISM). Grains of ISM origin 
were recently 
discovered in the inner solar system and at Earth orbit and may accrete
onto Earth after ablation in the atmosphere.
A favorable matrix for detection of such extraterrestrial material
is presented  by deep-sea sediments with very low sedimentation 
rates (0.8--3 mm/kyr).
We report here on the measurement of Pu isotopic abundances
in a 1-kg deep-sea dry sediment collected in 1992 in the North Pacific. 
Our estimate
of $(3\pm 3)\times 10^5$ $^{244}$Pu atoms in the Pu-separated fraction
of the sample shows no excess over the expected stratospheric nuclear
fallout content and under reasonable assumptions sets a limit of 
0.2 $^{244}$Pu atoms/cm$^2$yr for extra-terrestrial deposition.
Using available data on ISM steady-state flux
on Earth, we derive a limit of
$2\times 10^{-11}$ g-$^{244}$Pu/g-ISM for the abundance of $^{244}$Pu
in ISM.

\end{abstract}


\section{Introduction}
The legacy of short-lived nuclides, now extinct in the solar system,
is important
since their production spans a smaller number
of nucleosynthesis events than that of stable nuclides. Such events occur
continuously in the Galaxy and it has been estimated \citep{mey:00} that
radioactive species with lifetimes $\stackrel{>}{\sim}5-10$ Myr
should attain approximate steady-state abundances
in the interstellar medium (ISM).
The case of  $^{244}$Pu (t$_{1/2}$=81 Myr)
is interesting \citep{was:96},
as a pure {\it r-}process nuclide produced in supernova and neutron-star
disruption.
Its presence in early-solar material
has been inferred from Xe isotopic anomalies \citep{ros:65,ale:71,was:85}.
The  early-solar $^{244}$Pu/$^{238}$U ratio was
established to be $\sim$0.007 \citep{hud:89} which is about
the value  expected from uniform production \citep{was:96}.

The study of interstellar matter (ISM) has attracted considerable attention 
in the last decade, leading to the discovery of ISM grains in the
inner solar system by the Ulysses \citep{gru:93}
and Galileo space missions \citep{bag:95}. These grains whose masses
range from 10$^{-15}$ to above 10$^{-11}$ g are well identified as
ISM by  their retrograde trajectories, opposite to orbits of most
interplanetary grains and close to that of interstellar gas (26 kms$^{-1}$
from respectively 253$^\circ$ and 5$^\circ$ ecliptic longitude and latitude),
their constant flux in and out of the ecliptic plane and
their high speeds in excess of the solar-system escape speed
\citep{gru:00}.
Smaller-size particles are impeded from entering the planetary system
by electromagnetic interactions with the ambient magnetic field. 
The elemental and isotopic composition of these grains will be 
a major goal of research
with their recovery to Earth by the Stardust
mission \citep{bro:96} planned for 2006. 
The modeling of the existing body of data from Galileo and Ulysses
suggests that they can be matched only by assuming that there is no reduction 
of the interstellar dust flux at least as close as  1.3 AU and that up to 30\%
of the dust flux with masses above 10$^{-13}$ g at these distances is of
interstellar origin \citep{gru:97}.
More recently, even larger dust grains ($\sim 3\times 10^{-7}$ g)
incident on Earth and ablating in the atmosphere have been reliably 
identified as interstellar
by their hyperbolic speeds and directions with the
ground-based AMOR radar array \citep{bag:00,lan:00}.
An unmistakable signature for such
material accreting on Earth
would be the presence of short-lived nuclides now extinct in
solar--system  material.

We report on a search \citep{sak:99,val:00}
for $^{244}$Pu accumulated on Earth
through cosmic dust deposition. This search is based
on deep-sea sediment which, owing to very low sedimentation rates
($\sim$ 0.8--3 mm/kyr, see Goldberg and Koide 1962), offers a favorable matrix;
on the other hand the nuclide $^{244}$Pu 
can be detected with 
high sensitivity by accelerator mass spectrometry (AMS) \cite{fif:00}.
Evidence
for the presence of $^{244}$Pu in a rare-earth mineral 
was reported \citep{hof:71} but not confirmed since.
The detection of short-lived nuclides 
injected from a hypothetical 
near-Earth supernova
has also been considered 
\citep{ell:96,kni:99,fie:99} and evidence for $^{60}$Fe
in a deep-ocean ferromanganese crust,
interpreted as originating from such an event, has been
published \citep{kni:99}. The same group recently reported 
measurements of Pu isotopes in a manganese crust
\citep{wal:00}.

\section{$^{244}$Pu analysis of a deep-sea sediment sample}
  A deep-sea sediment sample was treated at 
Kanazawa University for Pu extraction and $\alpha$ counting
\citep{sak:99}. The sample (No. 92SAD01), originating 
from 80 kg of wet red clay,
was dredged in 1992 from 5,800 m in water depth in the
Pacific Ocean (9$^{\circ}$30'N, 174$^{\circ}$18'W) 
over a layer of seabed of $\sim$ 2 m$^2$ area and
$\stackrel{<}{\sim}0.3$ m sediment depth.
After washing and filtering off nodules,
an aliquot of 1.054 kg
of the resulting 12.85 kg dry sediment was used.
$^{239,240}$Pu of nuclear-bomb
fallout origin present in the sample
was used throughout as an in-situ tracer for the 
$\alpha$ and AMS analysis.
Pu was extracted by alkali fusion and anion exchange and 
electrodeposited on a stainless steel disk.
A 34 g aliquot of the dry-sediment sample was processed separately
after addition of a $^{242}$Pu tracer 
to determine the chemical efficiency
of Pu extraction ($\sim$40\%).
A value of
34$\pm$3 $\mu$Bq-$^{239+240}$Pu/g was determined for the dry sediment
from $\alpha$ counting of the latter sample and two independent $\alpha$
measurements.
A 546-day $\alpha$-counting  of the main Pu fraction was performed
using an $\alpha$ spectrometer with an efficiency of
$\sim$25\%. The dominant $\alpha$ group observed  corresponds to
$^{239+240}$Pu with 
an activity of
$\sim$0.013 Bq or $\sim$1.1$\times$10$^{10}$ $^{239+240}$Pu atoms.

   After $\alpha$-counting, the Pu-electrodeposited disk of the main fraction
was shipped to the Hebrew University for AMS analysis. The Pu
layer was dissolved in a HCl solution containing 2.4 mg Fe$^{3+}$ and 
Pu co-precipitated with iron (III)
hydroxide using ammonia solution. 
The overall efficiency of Pu recovery was determined to be $\sim$80\%.
The precipitate was dried, ignited to Fe$_2$O$_3$, divided
in two parts, then
pressed into the cathode holders
of a Cs-sputter negative ion source
\citep{gel:97}.
For normalization of the AMS analysis, calibration 
cathodes containing (3.2$\pm$0.3)$\times$10$^{10}$ $^{244}$Pu atoms were
prepared with the same procedure after spiking with
a $^{244}$Pu solution (0.32 ng-$^{244}$Pu/mL) prepared
at Argonne National Laboratory. The AMS analysis
was performed at the Koffler 14UD Pelletron tandem accelerator
(Weizmann Institute)
with the setup described in detail in Berkovits et al. 2000. 
$^A$Pu (A = 239,240,241,242,244)
were sequentially analyzed
by injecting $^A$PuO$^-$ into the accelerator. 
After
acceleration with
a terminal voltage of 7.1 MV, $^A$Pu$^{9+}$ ions were momentum and
velocity-analyzed before entering a detection system determining
time of flight and energy. Careful
calibration of the voltages of the accelerator and of the
Wien filter were required to analyze the 
Pu isotopes through the beam transport system, keeping magnetic rigidity
constant. Unambiguous identification of $^A$Pu ions 
was obtained in the detector from the time-of-flight and energy signals.

Fig. 1a shows the Pu isotopic distribution
measured by AMS for the calibration cathodes, in
good agreement with a distribution determined by $\alpha$ spectrometry.
The cathodes were
run to exhaustion of the Pu output ($\sim$ 4 hours), yielding
an overall detection efficiency (counts$/$atoms in cathode)
of $\sim$3$\times$10$^{-6}$.
The measured isotopic distribution
for the sediment sample is shown in fig. 1b.
The average count rate
for $^{239}$Pu over the lifetime of the cathode was 2.2 counts per
second.
The absolute number of combined $^{239+240}$Pu atoms, 
determined in the AMS measurement using
the above efficiency, 
is 1.4$\times$10$^{10}$, in reasonable
agreement with the $\alpha$-counting value.
The measured ratio $^{240}$Pu/$^{239}$Pu is 0.15$\pm$0.02,
in agreement with the stratospheric fallout value
0.18$\pm$0.01 (see Cooper et al 2000).
Reasonable agreement is also observed for $^{241}$Pu after correction for
its radioactive decay. The measured relative abundance of $^{242}$Pu
is in excess of the expected value for fallout or
any natural source and is attributed to a possible
contamination originating in the use of a $^{242}$Pu tracer
in the laboratory. The high isotopic purity of the tracer $(>90\%)$
precludes any significant effect on other Pu isotopes.
Special care was taken to 
reduce any memory effect in the AMS ion source
between the measurement with a calibration cathode 
and that of the sediment
sample. A run of 4 hours before the latter measurement, using a stainless
steel dummy cathode in the ion source, yielded no counts of $^{244}$Pu.
Only
one count of $^{244}$Pu with no background ions
was detected for the sediment sample during a 3.5-hour measurement, 
corresponding to $(3\pm3)\times 10^5$  $^{244}$Pu atoms in the cathode.
The $^{244}$Pu abundance in stratospheric nuclear fallout is not known 
and we estimate it here by extrapolating measured 
$^{A+1}$Pu/$^A$Pu (A = 239--241) fallout ratios (see Cooper et al. 2000). 
We use in this extrapolation 
the experimental distribution of Pu isotopes (A = 239--244) determined in
the Mike thermonuclear explosion \citep{dia:60} and assume that the
effective neutron capture cross sections in this experiment are the same as
for the global nuclear atmospheric tests. The double ratio
($^{244}$Pu/$^{242}$Pu)$_s$/($^{244}$Pu/$^{242}$Pu)$_M$ 
is then given by ($<$$\Phi$$>$$_s/$$<$$\Phi$$>$$_M$)$^2$, 
where $s$ and $M$ denote respectively
stratospheric and Mike tests and $<$$\Phi$$>$  
denotes the effective neutron fluence.
The  ratio $<$$\Phi$$>$$_s/$$<$$\Phi$$>$$_M$  
obtained from the data for $^{A+1}$Pu/$^A$Pu (A = 239--241)
is $0.52\pm 0.12$. The fallout--$^{244}$Pu content of 
our sample is thus estimated
as $(7\pm 3)\times 10^5$ atoms.
We conclude from a comparison with our measurement that any excess 
$^{244}$Pu accumulated in deep-sea
sediment from extraterrestrial sources is $<1\times 10^6$ 
$^{244}$Pu atoms/kg--dry sediment 
at a 90\% confidence level. 

\section{Limits on $^{244}$Pu abundance}
The fact that no $^{244}$Pu signal is observed above the nuclear-fallout
value is significant in that it helps to place  a bound on the $^{244}$Pu
abundance in the ISM.
We shall take $r_{sed}$ = 1 mm/kyr as a
representative value for the sedimentation rate in deep-sea pelagic
sediment \citep{gol:62}.
Based on the short residence times of Pu in the atmosphere and in the ocean
($<30$ yrs, Lal 2001), we can assume that the 
steady-state accumulation rate of ISM--$^{244}$Pu
in deep-sea sediment is equal to its 
accretion rate on Earth. Using the above limit for $^{244}$Pu 
concentration $C^{244}_{sed}$ in the sediment, 
a sediment density $\rho _{sed}$ = 1.5 g/cm$^3$, 
we obtain a limit for the steady-state interstellar $^{244}$Pu
deposition on Earth $\phi^{244}_{\oplus} =
 10^{-7} C^{244}_{sed}\rho _{sed}r_{sed} <
 0.2\;\; ^{244}$Pu atoms/cm$^2$ yr.

The ISM accretion flux onto Earth is highly uncertain at present.
A quantitative estimate may however be attempted using the recent
experimental AMOR data \citep{bag:00,lan:00} which determine 
a flux $\phi ^{ISM}_{1AU}$ of $\sim 1.8\times 10^{-12}$
and of $<3\times 10^{-14}$ cm$^{-2}$s$^{-1}$ respectively at southern
and northern ecliptic latitudes, for ISM grains of
mass $m^{ISM}_{grain}>3\times 10^{-7}$ g. These values 
roughly fit, with a $m^{-1}$  
power law, the mass distribution of the interstellar grains 
identified by the Ulysses mission \citep{lan:00}.  
The flux $\phi ^{ISM}_{1AU}$ is highly
anisotropic and, assuming that unidirectional particles ablating 
on the Earth's cross section average homogeneously over the Earth's
surface, we obtain an ISM steady-state deposition flux
$\phi^{ISM}_{\oplus}= \phi^{ISM}_{1AU}m^{ISM}_{grain}/4
   >1.4\times 10^{-19}$ g-ISM/cm$^2$s.
The corresponding limit on the abundance of $^{244}$Pu in ISM is
\begin{equation}
  C^{244}_{ISM}=
  3.2\times 10^{-8}(\phi^{244}_{\oplus} \;244/{\cal N_A})/
  \phi^{ISM}_{\oplus}
  <2\times 10^{-11}\;\textrm{g-}^{244}\textrm{Pu/g-ISM}.
  \label{eq:CISM}
\end{equation}
Further assuming that ISM has an early solar-system  uranium
abundance of $1.6\times 10^{-8}$ g-U/g \citep{and:89},
we derive a limit of $(^{244}Pu/U)_{ISM} < 1\times 10^{-3}$
in steady-state ISM.
This upper limit is smaller by about one order of magnitude 
than the
adopted early-solar ratio $(^{244}Pu/U)_\odot =0.007$. 
The latter value was shown to be 
about half of that calculated with a uniform 
production (UP) model \citep{was:96} of actinides by supernova 
{\it r-}process and it may be significant that the 
($^{244}Pu/U)_{ISM}$ limit estimated here lies 
even lower. 
Wallner et al. 2000 recently reported on a $^{244}$Pu measurement in
a 120 g ferromanganese crust where they detected one count of $^{244}$Pu
corresponding
to $\sim 1\times 10^5$ $^{244}$Pu atoms. The authors derive from a known 
accumulation age of the crust ($\sim 13$ Myr),
a flux of $\sim 300\;^{244}$Pu atoms/cm$^2$Myr and conclude that
the ratio between this flux and the flux of $^{60}$Fe measured in a
similar deep-ocean crust is compatible with the
signal of a near-Earth ($\sim 30$ pc) supernova explosion $\sim 5$ Myr ago
\citep{kni:99}.
It is interesting to interpret this $^{244}$Pu flux estimate in
terms of steady-state ISM deposition rather than effects of 
ejecta from a near-Earth supernova.
Using the assumptions and 
estimates detailed above 
for the ISM steady-state accretion flux onto Earth, 
the flux determined by Wallner et al. 
would lead to a ratio 
($^{244}Pu/U)_{ISM}\;\stackrel{\textstyle <}{\sim}2\times 10^{-6}$, smaller
by  three orders of magnitude
than the early-solar value or the UP model. We do not know 
at this point whether the low ($^{244}$Pu/U)$_{ISM}$ values
estimated in the present work
reflect 
({\it i}) an overestimate of the steady-state ISM flux onto Earth, 
({\it ii}) a possible
fractionation process reducing the Pu transport to Earth or,
for the ferromanganese crust, reducing the Pu intake
or ({\it iii}) perhaps, represent a genuine low $^{244}$Pu
abundance in ISM.

The experimental limits derived in the present work were
constrained by the nuclear-fallout Pu content
of the deep-sea sediment sample. This ensured
that Pu geochemical and chemical behaviors were under control; however,
one could establish more stringent limits by using
a deep-sea core sample 
that is clean of artificial Pu input and by improving the
overall sensitivity of the $^{244}$Pu detection.
We are presently
pursuing these directions.

We should like to thank Y. Kashiv (U. of Chicago) for useful
discussions and A. Kaufmann (Weizmann Institute) for his help.
The authors are indebted for the use of the $^{244}$Pu
calibration sample to the Office of the
Basic Energy Sciences, U.S. Department of Energy, through the
transplutonium element production facilities at Oak Ridge National Laboratory.
This work was supported in part by the German Israel Foundation-GIF (HU)
and by the US-D.O.E. Nuclear Science Division, under Contract No.
W-31-109-ENG-38 (ANL).

\newpage
\begin{figure}
\begin{center} 
\epsfig{file=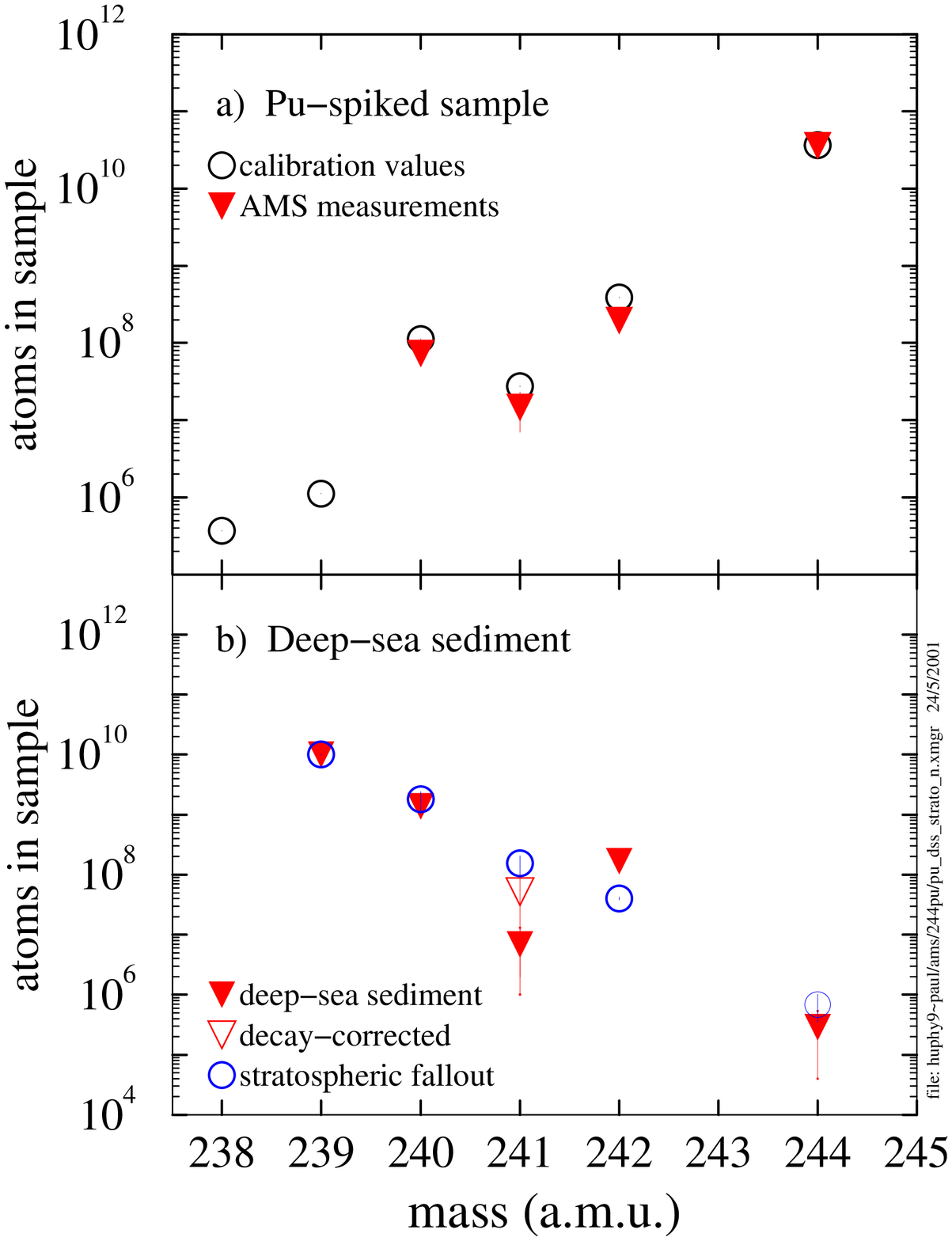,width=11.5cm}
\caption{Isotopic distribution of plutonium (solid triangles) 
measured by accelerator mass 
spectrometry (AMS) for : {\it a)} 
a calibration sample containing $3.7\times 10^{10}$
$^{244}$Pu atoms. The open circles show the distribution measured by 
$\alpha$-counting of a thin source prepared from the same solution as the
AMS calibration sample. The AMS results were normalized to the
number of $^{244}$Pu atoms obtained from the $\alpha$ counting.
{\it b)} the plutonium fraction extracted from the 1.020-kg deep-sea
dry sediment 92SDA01. The AMS results were normalized
to the combined number of $^{239,240}$Pu atoms measured by $\alpha$ 
spectrometry. The open circles represent the
expected Pu isotopic distribution of nuclear-bomb stratospheric fallout.
The open triangle for A=241 includes the
radioactive decay correction of fallout $^{241}$Pu. 
See text for details.}
\label{fig.1}
\end{center}
\end{figure}


\end{document}